\PassOptionsToPackage{prologue,dvipsnames}{xcolor}
\documentclass[aps,pre,twocolumn,showpacs,groupedaddress,superscriptaddress,footinbib,longbibliography]{revtex4-2}

\usepackage{mathrsfs}
\usepackage{hyperref}
\setcitestyle{square,sort&compress,comma,numbers}
\hypersetup{colorlinks=true, citecolor=blue, urlcolor=blue, linkcolor=blue}
\usepackage[english]{babel}
\usepackage[utf8]{inputenc}
\usepackage{graphicx}
\usepackage{amsmath}
\usepackage{amsfonts}
\usepackage{amssymb}
\usepackage{color,soul}
\setulcolor{red}
\usepackage{easyReview}
\usepackage{xcolor}
\usepackage{float}
\usepackage{newtxtext,newtxmath}

\usepackage[normalem]{ulem}

\newcommand\diff{\mathrm{d}}
\renewcommand{\vec}[1]{\boldsymbol{#1}}

\begin{document}
\title{Optimal first-passage times of active Brownian particles under stochastic resetting}

\author{Yanis Baouche}
\email{baouche@pks.mpg.de}
\affiliation{Max-Planck-Institut f{\"u}r Physik komplexer Systeme, N{\"o}thnitzer Stra{\ss}e 38,
01187 Dresden, Germany}
\author{Christina Kurzthaler}
\email{ckurzthaler@pks.mpg.de}
\affiliation{Max-Planck-Institut f{\"u}r Physik komplexer Systeme, N{\"o}thnitzer Stra{\ss}e 38,
01187 Dresden, Germany}
\affiliation{Center for Systems Biology Dresden, Pfotenhauerstra{\ss}e 108, 01307 Dresden, Germany}
\affiliation{Cluster of Excellence, Physics of Life, TU Dresden, Arnoldstra{\ss}e 18, 01062 Dresden, Germany}

\begin{abstract} 
We study the first-passage-time (FPT) properties of an active Brownian particle under stochastic resetting to its initial configuration, comprising its position and orientation, to reach an absorbing wall in two dimensions. Coupling a perturbative approach for low P{\'e}clet numbers, measuring the relative importance of self-propulsion with respect to diffusion, with the renewal framework for the stochastic resetting process, we derive analytical expressions for the survival probability, the FPT probability density, and the associated low-order moments. Depending on their initial orientation, the minimal mean FPT for active particles to reach the boundary can both decrease and increase relative to the passive counterpart. The associated optimal resetting rates depend non-trivially on the initial distance to the boundary due to the intricate interplay of resetting, rotational Brownian noise, and active motion. 
\end{abstract} 

\maketitle

\section{Introduction}

Stochastic resetting is a relatively recent concept whereby a process is randomly reset to a predetermined state. Mainly introduced in the context of search processes with Brownian motion \cite{evansDiffusionStochasticResetting2011c, whitehouseEffectPartialAbsorption2013}, this framework has been widely expanded to take into account other types of processes, such as L{\'e}vy flights \cite{kusmierzFirstOrderTransition2014, kusmierzOptimalFirstarrivalTimes2015, radiceOptimizingLeapoverLengths2024}, non-Poissonian waiting times \cite{euleNonequilibriumSteadyStates2016, shkilevContinuoustimeRandomWalk2017, kurzthalerCharacterizationControlRunandTumble2024a, zhaoQuantitativeCharacterizationRunandtumble2024a}, or partial \cite{tal-friedmanDiffusionPartialResetting2022, biroliResettingRescalingExact2024}, time-dependent \cite{palDiffusionTimedependentResetting2016}, or random resetting mechanisms~\cite{mendezFirstpassageTimeBrownian2024}. Overall, stochastic resetting has proven to be a very fruitful research direction for the physics community with a variety of applications. The most common applications being algorithmics \cite{lubyOptimalSpeedupVegas1993, montanariOptimizingSearchesRare2002,blumerStochasticResettingEnhanced2022,blumerCombiningStochasticResetting2024}, chemical reactions \cite{coppeyKineticsTargetSite2004,chmeliovFluorescenceBlinkingSingle2013,ghoshFirstpassageProcessesFilamentous2018}, and animal foraging~\cite{benichouOptimalSearchStrategies2005, benichouIntermittentSearchStrategies2011}. In all of those applications, resetting prevents being trapped in a suboptimal state (e.g., resource- or reactant-depleted zone), and most importantly, expedites the search completion (food sources, reactants or optima), thus improving performance when time constraints are imposed.

The question of FPT properties is particularly important in the context of biology, where the efficiency during foraging is a crucial part of a microorganism's ability to survive and achieve its biological purpose \cite{suarezSpermTransportFemale2006, laugaHydrodynamicsSwimmingMicroorganisms2009, viswanathanPhysicsForagingIntroduction2011, bechingerActiveParticlesComplex2016,laugaFluidDynamicsCell2020, kurzthalerOutofequilibriumSoftMatter2023, grebenkovTargetSearchProblems2024}. Using biology as a model system, active agents have been engineered in the lab and extract energy from their environment to self-propel \cite{ebbensPursuitPropulsionNanoscale2010,barabanTransportCargoCatalytic2011,gaoHydrogenBubblePropelledZincBasedMicrorockets2012,ebbensActiveColloidsProgress2016,robinSinglemoleculeTheoryEnzymatic2018}. Establishing a physical understanding of how fast these active colloids reach specific targets is an important aspect in their design for target-delivery and bioremediation applications~\cite{azizTrackingMagneticMicromotors2021a,azizImagingControlNanomaterialdecorated2022a,nauberMedicalMicrorobotsReproductive2023a}. Because of the coupling between translational and rotational degrees of freedom, studying the FPT properties of active agents is arguably more difficult than the passive (Brownian) counterpart, but can also lead to a richer variety of results~\cite{Angelani:2014, Malakar:2018, tucciFirstpassageTimeRunandtumble2022,vinodNonergodicityResetGeometric2022, ditrapaniActiveBrownianParticles2023b, gueneauOptimalMeanFirstpassage2024,gueneauRunandtumbleParticleOnedimensional2025}. In this paper, we study one of the paradigmatic models of active matter -- the active Brownian particle (ABP) -- comprising the effect of active motion, translational and rotational diffusion \cite{howseSelfMotileColloidalParticles2007b,romanczukActiveBrownianParticles2012a,tenhagenBrownianMotionSelfpropelled2011a,kurzthalerIntermediateScatteringFunction2016a,kurzthalerProbingSpatiotemporalDynamics2018b}.

While it is well-established that resetting a passive tracer to its original position enables a stationary distribution and a finite mean FPT for reaching the target \cite{masoliverTelegraphicProcessesStochastic2019,evansStochasticResettingApplications2020a,evansDiffusionStochasticResetting2011c}, the interplay of resetting with active motion has not yet been studied in the context of FPT statistics. Building upon our recent findings for the ABP at low P{\'e}clet numbers~\cite{baoucheFirstpassagetimeStatisticsActive2025}, where we determined analytical predictions for the survival distribution and FPT probability density, and combining it with the typical renewal approach used for stochastic resetting~\cite{evansStochasticResettingApplications2020a}, we study the interplay of resetting and self-propulsion. In particular, at the resetting event the agent is reset to its initial configuration, comprising its position and orientation. Our main results suggest that, owing to the strong dependence on the initial orientation, resetting in active motion can either increase of decrease the mean MFPT to reach the absorbing boundary in comparison to the passive case. Notably, we measure this directional bias through an anisotropy function and find that it becomes most pronounced when the initial distance to the wall is comparable to the distance traveled by the agent before being reset.

This paper is organized as follows: In Sec.~\ref{sec:model} we first introduce our model. We then summarize the perturbative approach used to study the FPT statistics of an ABP without stochastic resetting and, using a renewal approach, we link the survival probability with and without stochastic resetting. This framework allows us to analytically compute various statistical indicators, including the mean FPT, the median, and the skewness, and we finally resolve the full survival probability and FPT probability density, which we present in Sec.~\ref{sec:results}. We summarize and conclude in Sec.~\ref{sec:summary}.

\begin{figure}[ht]
\includegraphics[width =0.8\linewidth, scale=0.05]
{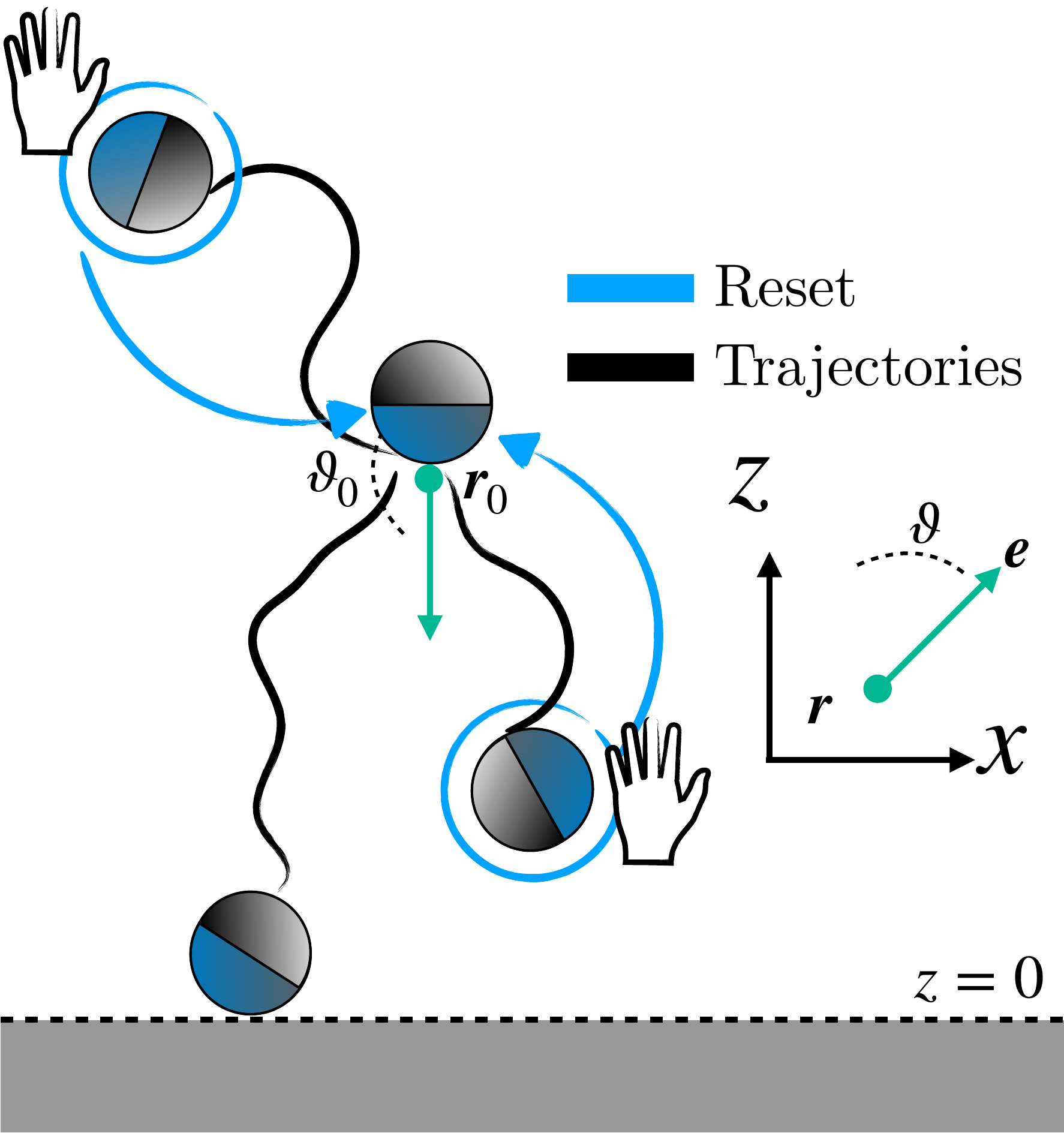}
\caption{Schematic of the motion of an ABP under stochastic resetting to its initial position $\vec{r}_0$ and orientation $\vartheta_0$  near an absorbing boundary at $z=0$. The inset depicts the agent's position $\vec{r}$ and orientation $\vartheta$. \label{fig:schematic}}
\end{figure}

\section{Model \label{sec:model}}
In this section, we outline the strategy employed to obtain the first-passage-time (FPT) statistics of an active Brownian particle (ABP), whose position and orientation are reset to their initial state at random times.
We consider an ABP moving in a two-dimensional (2D) plane~$(O, x, z)$. The particle moves at a constant speed~$v$ along its instantaneous orientation~$\vec{e}(\vartheta(t)) =(\sin(\vartheta(t)), \cos(\vartheta(t)))$, where $\vartheta(t)$ denotes the polar angle [Fig.~\ref{fig:schematic}({\it inset})]. The agent undergoes both random translational and rotational motion, characterized by their respective diffusion coefficients $D$ and $D_{\mathrm{rot}}$. In addition, the instantaneous position~$\vec{r}(t)$ and orientation~$\vartheta(t)$ are randomly reset to the initial position~$\vec{r}_{0}\equiv \vec{r}(0)$ and orientation~$\vartheta_{0} \equiv \vartheta(0)$; the resetting events happen at times drawn from an exponential distribution $\phi(t)=\lambda \exp(-\lambda t)$ with rate~$\lambda$ (and mean time between resets~$1/\lambda$) (see Fig.~\ref{fig:schematic}). These processes are represented by the following set of stochastic equations:
\begin{subequations}
\begin{align}
    \frac{\diff \vec{r}}{\diff t} &= v \vec{e} + \sqrt{2D} \boldsymbol{\eta},\label{eq:stoch_r} \\ 
    \frac{\diff \vartheta}{\diff t} &= \sqrt{2D_{\mathrm{rot}}} \xi, \label{eq:stoch_theta}\\
    \vec{r} &\to \vec{r}_{0} \text{ and } \vartheta \to \vartheta_{0} \text{ with } T(t), 
\end{align}
\end{subequations}
where $\vec{\eta}(t)$ and $\xi(t)$ are independent Gaussian white noises of zero mean and delta-correlated variance, $\langle \eta_{i}(t) \eta_{j}(t') \rangle = \delta_{ij}\delta(t-t')$ for $i,j \in \{1,2 \}$ and $\langle \xi(t) \xi(t') \rangle = \delta(t-t')$. Rescaling the position $\vec{r} =a \vec{\mathcal{R}}$ with the particle's hydrodynamic radius $a$ and the time $t=\tau T$ with the diffusive time $\tau=a^{2}/D$, two non-dimensional numbers appear: $\mathrm{Pe}=va/D$ denotes the P{\'e}clet number, measuring the relative importance of active motion versus diffusion, and $\Lambda=\lambda \tau$ is the reduced resetting rate, representing the prevalence of resetting events compared to the diffusive time scale. We further introduce $\gamma=\tau D_{\mathrm{rot}}=3/4$, which follows from the Stokes-Einstein-Sutherland relation for a spherical particle. 

\subsection{Renewal framework \label{sec:renewal}}
To make analytical progress, we rely on a renewal approach for the survival probability $S(T|Z_0,\vartheta_0)$, i.e. the probability that the agent, which has started at position~$Z_0$ with orientation $\vartheta_0$, has not yet reached the boundary at time $T$. For exponentially distributed resetting times $\phi(T)=\Lambda\exp(-\Lambda T)$, the survival probability  follows~\cite{evansStochasticResettingApplications2020a}:
\begin{align}
\begin{split}
    &S(T| Z_{0}, \vartheta_{0}) = e^{-\Lambda T}S^{\mathrm{o}}(T |Z_{0}, \vartheta_{0}) \\
    &+ \int_{0}^{T} \Lambda e^{-\Lambda T}S^{\mathrm{o}}(T'|Z_{0}, \vartheta_{0}) S(T-T'|Z_{0}, \vartheta_{0})~\mathrm{d}T', 
\end{split}\label{eq:renewal}
\end{align}
where $S^{\mathrm{o}}(T'|Z_{0}, \vartheta_{0})$ corresponds to the survival probability of an ABP in the absence of resetting. Equation~\eqref{eq:renewal} is to be interpreted in the following way: the probability to survive up to time~$T$ is the sum of the probability to survive up to time~$T$ without any resetting event and the sum over the probabilities to survive up to time $T'$ given a reset at an earlier time $T-T'$. The survival probability provides access to the FPT probability density $F(T| Z_{0}, \vartheta_{0})$, characterizing the distribution of times at which the agent reaches the wall:
\begin{equation}
F({T} | {Z}_{0}, \vartheta_{0})= - \frac{\diff }{\diff {T}}S({T} | {Z}_{0}, \vartheta_{0}). \label{eq:fpt_derivative_survival}
\end{equation}

To compute the survival probability we employ a Laplace transform $T\mapsto s$
\begin{align}
\widehat{S}(s|Z_0, \vartheta_0) = \int_{0}^{\infty}   S(T|Z_0, \vartheta_0) e^{-sT}~\mathrm{d}T,
\end{align}
which allows us to readily obtain a closed-form solution 
\begin{equation}
    \widehat{S}(s|Z_{0}, \vartheta_{0}) = \frac{\widehat{S}^{\mathrm{o}}(s+\Lambda,Z_{0}, \vartheta_{0}) }{1 - \Lambda \widehat{S}^{\mathrm{o}}(s+\Lambda,Z_{0}, \vartheta_{0})}. \label{eq:survival_reset}
\end{equation}
It then suffices to know $\widehat{S}^{o}$ -- the survival probability without resetting -- to obtain the survival probability with resetting $\widehat{S}$ in Laplace space. 

The former quantity was the object of our previous work~\cite{baoucheFirstpassagetimeStatisticsActive2025}, where we have derived analytical expressions for $S^{\mathrm{o}}$ through a perturbation expansion for small P{\'e}clet numbers. While details of the perturbation expansion can be found in Ref.~\cite{baoucheFirstpassagetimeStatisticsActive2025} and Appendix~\ref{appendix:perturbative}, we recapitulate the most important steps here. 
Starting from Eqs.~\eqref{eq:stoch_r}-\eqref{eq:stoch_theta}, we first derive a (non-dimensional) Fokker-Planck equation for the probability density $\mathbb{P}^{\mathrm{o}}(\vec{\mathcal{R}}, \vartheta,T | \vec{\mathcal{R}}_{0}, \vartheta_{0})$ of a particle to be at~$\vec{\mathcal{R}}$ with orientation~$\vartheta$ at time $T$ having started at~$\vec{\mathcal{R}}_{0}$ with orientation~$\vartheta_0$ at $T=0$:
\begin{align}
    \partial_{T} \mathbb{P}^{\mathrm{o}} &= - \mathrm{Pe}~ \vec{e} \cdot \boldsymbol{\nabla} \mathbb{P}^{\mathrm{o}} +  \gamma \partial_{\vartheta}^{2} \mathbb{P}^{\mathrm{o}} +  \nabla^{2} \mathbb{P}^{\mathrm{o}}. \label{eq:FPE_tot} 
\end{align}
Since the boundary is infinite in the $X$ direction, we further integrate out the $X$ component and arrive at
\begin{align}
    \partial_{{T}} \mathbb{P}^{\mathrm{o}} &= - {\mathrm{Pe}} \cos(\vartheta) \partial_{Z} \mathbb{P}^{\mathrm{o}} +\gamma \partial_{\vartheta}^{2}  \mathbb{P}^{\mathrm{o}} +\partial_{{Z}}^2\mathbb{P}^{\mathrm{o}}, \label{eq:FP}
\end{align}
which is supplemented by the following initial and boundary conditions:
\begin{subequations}
\begin{align}
    &\mathbb{P}^{\mathrm{o}}(Z, \vartheta, T=0 | Z_{0}, \vartheta_{0}) = \delta(Z-Z_{0}) \delta(\vartheta - \vartheta_{0}),\\
    &\mathbb{P^{\mathrm{o}}}(Z=0, \vartheta,T| Z_{0},\vartheta_{0})  = 0 \quad  \forall T \in \mathbb{R}^+. \label{eq:BC}
\end{align}
\end{subequations}
Next, we move to Laplace space ($s\mapsto T$) and thus Eq.~\eqref{eq:FP} transforms to 
\begin{equation}
    (s-\mathcal{H})\widehat{\mathbb{P}}^{\mathrm{o}} = \delta(Z-Z_{0})\delta(\vartheta-\vartheta_{0}), \label{eq:operator_form}
\end{equation}
where the operator $\mathcal{H} \equiv \mathcal{H}_{0} + \mathrm{Pe}\mathcal{V}$ is split into two components: the unperturbed operator $\mathcal{H}_0\equiv \gamma \partial_{\vartheta}^{2}  +\partial_{{Z}}^2$ and the perturbation $\mathcal{V} \equiv -\cos(\vartheta) \partial_{Z}$. Expanding the probability density in terms of the P{\'e}clet number 
\begin{align}
\widehat{\mathbb{P}}^{\mathrm{o}} = \widehat{\mathbb{P}}^{\mathrm{o}}_{0} + \mathrm{Pe} \ \widehat{\mathbb{P}}_{1}^{\mathrm{o}} + \mathrm{Pe}^{2} \ \widehat{\mathbb{P}}_{2}^{\mathrm{o}} + \mathcal{O}(\mathrm{Pe}^{3}), \label{eq:phat}
\end{align}
and inserting it into Eq.~\eqref{eq:operator_form}, leads to a set of coupled equations for the perturbations [Eqs.~\eqref{eq:G}-\eqref{perturb_eq2} in Appendix~\ref{appendix:perturbative}]. Analytical expressions for $\widehat{\mathbb{P}}_{k}^{\mathrm{o}}$ can be obtained iteratively and used as input for the survival probability $\widehat{S}^{\mathrm{o}}(s|Z_{0}, \vartheta_{0}) = \sum_{n=0}^{\infty} {\mathrm{Pe}}^{n}\widehat{S}_{n}^{\mathrm{o}}(s|Z_{0}, \vartheta_{0})$ via
the relation 
\begin{equation}
\widehat{S}^{\mathrm{o}}_k(s | {Z}_{0}, \vartheta_{0} ) = \int_{0}^{\infty}          \widehat{\mathbb{P}}^{\mathrm{o}}_k(s ,Z | Z_{0}, \vartheta_{0})~\mathrm{d}Z. \label{eq:survival_ABP}
\end{equation} 
Analytical predictions up to second order in the $\mathrm{Pe}$ number can be found in the Appendix~\ref{appendix:perturbative}. Finally, inserting the small-$\mathrm{Pe}$ expansion of $\widehat{S}^{\mathrm{o}}(s|Z_{0}, \vartheta_{0})$ into Eq.~\eqref{eq:survival_reset} yields the survival probability in the presence of stochastic positional resetting $\widehat{S}(s|Z_{0}, \vartheta_{0})$ and provides immediate access to the FPT probability density. 

To further quantify the FPT properties, we are interested in the low-order moments of the first-passage times, which are accessible through the survival probability in Laplace space $\widehat{S}$. In particular, the $n$-th moment of the random variable $F_{T}$ associated with the FPT probability density can be obtained via
\begin{subequations}
\begin{align}
    \mathbb{E}[F_{T}^{n}] &= \int_{0}^{\infty} T^{n} F(T|Z_0,\vartheta_0)~\mathrm{d}T, \\
    &= -\int_{0}^{\infty} T^{n} \left(\frac{\diff }{\diff T}S(T|Z_0,\vartheta_0) \right) e^{-sT}|_{s=0}~\mathrm{d}T, \\ 
    &=(-1)^{n+1} \frac{\diff^{n}}{\diff s^{n}} \left[s \widehat{S}(s|Z_0,\vartheta_0) \right]_{s=0} \label{eq:moments_survival},
\end{align}
\end{subequations}
where we have used the relation between $F(T|Z_0,\vartheta_0)$ and $S(T|Z_0,\vartheta_0)$ [Eq.~\eqref{eq:fpt_derivative_survival}] and the properties of the Laplace transform.

\subsection{Expansion in the P{\'e}clet number}
Note that we can further expand the survival probability for small $\mathrm{Pe}$, $\widehat{S}(s|Z_{0}, \vartheta_{0})=\sum_{n=0}^{\infty} {\mathrm{Pe}}^{n}\widehat{S}_{n}(s|Z_{0}, \vartheta_{0})$ [Eq.~\eqref{eq:survival_reset}] and formally obtain the associated coefficients:
\begin{subequations}
\begin{align}
\widehat{S}_{0} &= \frac{\widehat{S}_{0}^{\mathrm{o}}}{1-\Lambda \widehat{S}_{0}^{\mathrm{o}}},\\
\widehat{S}_{1} &=\frac{\widehat{S}_{1}^{\mathrm{o}}}{1-\Lambda \widehat{S}_{0}^{\mathrm{o}}}+ \frac{\Lambda\widehat{S}_{0}^{\mathrm{o}}\widehat{S}_{1}^{\mathrm{o}}}{\left(1-\Lambda \widehat{S}_{0}^{\mathrm{o}}\right)^2},\\
\widehat{S}_{2} &=\frac{\widehat{S}_{2}^{\mathrm{o}}}{1-\Lambda \widehat{S}_{0}^{\mathrm{o}}}+ \frac{\Lambda \left(\widehat{S}_{0}^{\mathrm{o}}\widehat{S}_{2}^{\mathrm{o}}+\left(\widehat{S}_{1}^{\mathrm{o}}\right)^2\right)}{\left(1-\Lambda \widehat{S}_{0}^{\mathrm{o}}\right)^2}+\frac{\Lambda^{2} \widehat{S}_{0}^{\mathrm{o}}\left(\widehat{S}_{1}^{\mathrm{o}}\right)^2}{\left(1-\Lambda \widehat{S}_{0}^{\mathrm{o}}\right)^3}, 
\end{align}
\end{subequations}
which can be readily extended to higher orders. Using this result, the FPT probability density in Laplace space obeys
\begin{align}
\begin{split}
\widehat{F}(s &| {Z}_{0}, \vartheta_{0})=\\
&= 1-s\widehat{S}_{0}(s|Z_{0}, \vartheta_{0}) -s\sum_{n=1}^{\infty} {\mathrm{Pe}}^{n}\widehat{S}_{n}(s|Z_{0}, \vartheta_{0}), \label{eq:fpt_laplace}
\end{split}
\end{align}
where the first two terms correspond to the FPT probability density of a Brownian particle under stochastic resetting and the sum encodes the effect of activity.  

\subsection{Brownian particle under stochastic resetting}
Our framework allows recovering the well-established result for the survival probability of a passive Brownian particle under stochastic resetting as
\begin{align}
\widehat{S}_{0}(s|Z_{0},\vartheta_{0})=\frac{1-e^{-\sqrt{\Lambda+s}Z_0}}{s+ \Lambda e^{-\sqrt{\Lambda+s}Z_0  }}. \label{eq:survival_Brownian}
\end{align}
Notably, we mentioned that introducing a resetting mechanism for a diffusive process establishes a finite mean first-passage time (MFPT). It reduces to 
\begin{align}
    \mathbb{E}[F_{T}]_{B} &= \widehat{S}_0(s=0) = \frac{e^{\sqrt{\Lambda}Z_{0}}-1}{\Lambda}, \label{eq:MFPT_B}
\end{align}
which represents the well-known MFPT for a diffusive particle under stochastic positional resetting~\cite{evansDiffusionStochasticResetting2011c}. It diverges as~$\propto \Lambda^{-1/2}$ for $\Lambda \to 0$, thus approaching the behavior of a Brownian particle without resetting. Furthermore, it diverges as $\Lambda \to \infty$, reflecting the particles that are constantly reset and never manage to reach the wall. Most importantly, for a fixed initial distance $Z_{0}$, there exists an optimal resetting rate $\Lambda^{*}_B = (Z^{*})^{2}/Z_{0}^{2}$ that minimizes the MFPT, where $Z^{*} = 1.59362...$ is the unique solution of the transcendental equation:
\begin{equation}
    \frac{Z^{*}}{2} = 1-e^{-Z^{*}}. \label{eq:transcendental_brownian}
\end{equation}
This indicates that resets far away from the wall should happen at small rates to minimize the MFPT. 

\begin{figure*}[tp]
\includegraphics[width =\textwidth]
{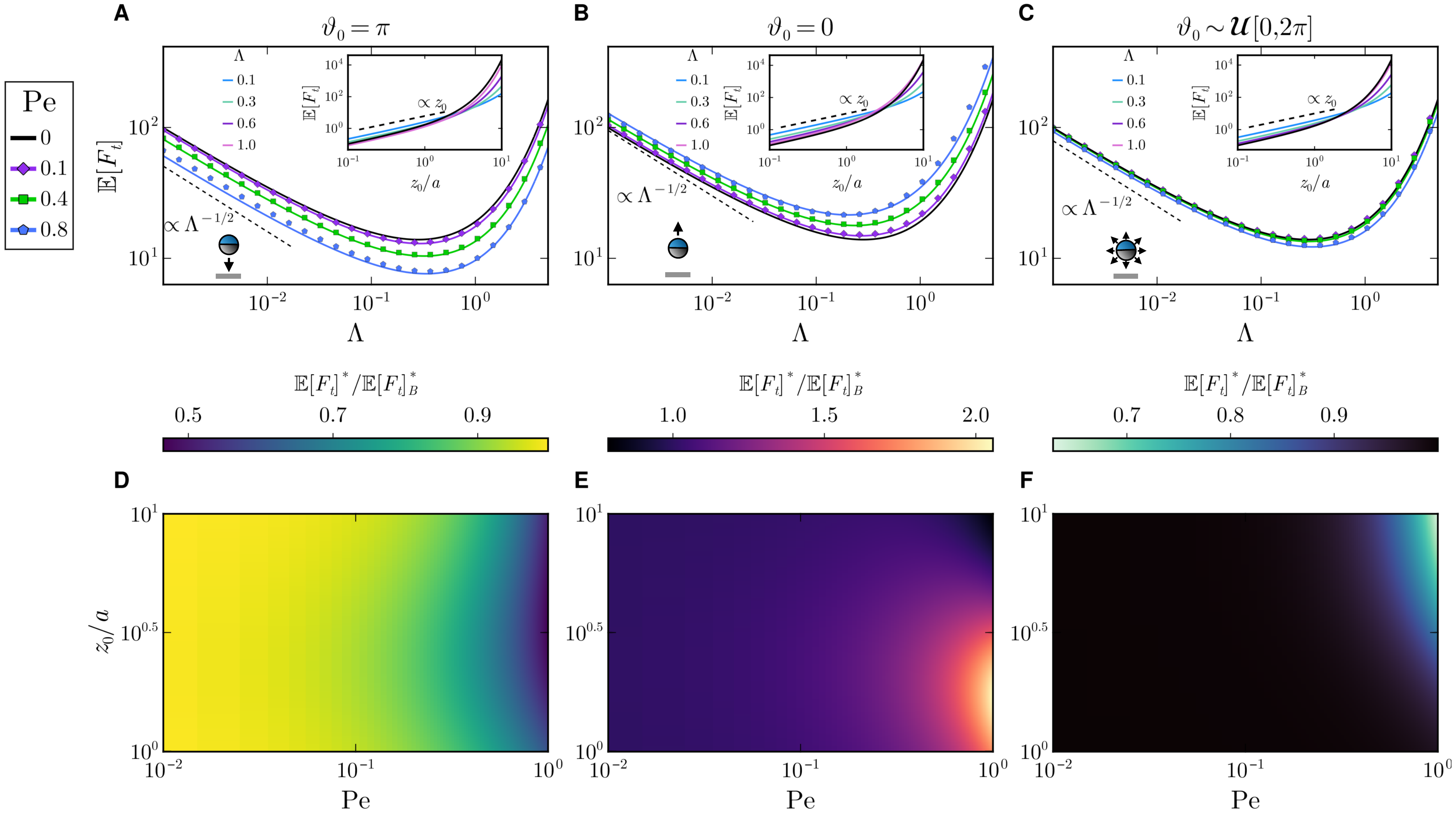}
\caption{(\textbf{A-C}) Mean first-passage time $\mathbb{E}[F_{t}]$ for three different initial angles $\vartheta_{0}$ as a function of the resetting rate $\Lambda$ (with $z_0/a=3$) for different P{\'e}clet numbers $\mathrm{Pe}$. Insets show the MFPT as a function of the initial distance $z_{0}$ for several $\Lambda$ and $\mathrm{Pe}=0.5$. ({\textbf{D-F}}) Ratio of the mean first-passage times of an active and passive Brownian particle at the optimal resetting rate $\mathbb{E}[F_{t}]^{*}/\mathbb{E}[F_{t}]^{*}_{B}$ as a function of $\mathrm{Pe}$ and $z_{0}$. Columns show results for particles (\textbf{A,D}) initially facing the wall, (\textbf{B,E}) initially facing against the wall, and (\textbf{C,F}) with initial angles drawn from a uniform distribution $\mathcal{U}[0, 2\pi]$. Solid lines and markers denote theory and simulations, respectively. The black lines in (\textbf{A-C}) correspond to the passive case with resetting rate $\Lambda_{B}=1$. \label{fig:mfpt_optimum}}
\end{figure*}

\section{Results \label{sec:results}}
In what follows, we show our results for the active Brownian particle under stochastic resetting up to the second order in the P{\'e}clet number. We discuss the mean first-passage times  [Sec.~\ref{sec:MFPT}], the anisotropy of the process [Sec.~\ref{sec:anisotropy}], the survival probability and the probability density for the FPT [Sec.~\ref{sec:FPTresults}], as well as the median and the skewness of the FPTs [Secs.~\ref{sec:median} and \ref{sec:skewness}].

\subsection{Mean first-passage time \label{sec:MFPT}}
We  compute the mean first-passage time (MFPT) by following the strategy outlined in Sec.~\ref{sec:renewal}. Our theoretical prediction reveals that it becomes finite [Eq.~\eqref{eq:MFPT} in Appendix \ref{appendix:results}] within the limit of small $\mathrm{Pe}$ and depends strongly on the initial orientation $\vartheta_{0}$ and initial distance to the boundary~$z_0$. Let us first note that Figs.~\ref{fig:mfpt_optimum} (\textbf{A}-\textbf{C}) show that  our theoretical predictions are nicely corroborated by simulation results (see Appendix~\ref{app:simulations} for details). We further note that though our results may loose accuracy as $\mathrm{Pe} \to 1$, we show in Appendix~\ref{appendix:validation_peclet} that our perturbation approach remains valid for this parameter range and we observe very good agreement with simulations even at larger P{\'e}clet numbers.

Importantly, we find that an agent initially facing the wall ($\vartheta=\pi$) reaches it faster than an agent oriented away from the wall ($\vartheta=0$), which is in turn slower than a diffusive agent [Figs.~\ref{fig:mfpt_optimum} (\textbf{A}-\textbf{B})]. Assuming uniformly distributed initial angles leads to a slightly lower MFPT than the diffusive case [Fig.~\ref{fig:mfpt_optimum} (\textbf{C})].  Additionally, increasing activity through the P{\'e}clet  number expedites and delays the arrival at the boundary for $\vartheta_{0}=\pi$ and $\vartheta_{0}=0$, respectively. This is explained by the fact that the distance the particle travels before reorienting (i.e., the persistence length) increases conjointly with activity, $l_{p}=v/D_\mathrm{rot}= a \mathrm{Pe}/\gamma$, and that the agent's initial orientation is also reset in the process. Hence, a particle initially departing from (resp. moving towards) the wall at a higher velocity will reach the wall at later (resp. earlier) times even with resetting. Importantly, the divergence of the MFPT $\propto\Lambda^{-1/2}$ as $\Lambda \to 0$ and for $\Lambda \to \infty$ is preserved for an active particle, as resetting too often prevents ever reaching the wall and not resetting enough leads to the divergence as for a simple ABP. 

The MFPTs further depend on the agent's initial distance~$z_{0}$ [Figs.~\ref{fig:mfpt_optimum} (\textbf{A}-\textbf{C})({\it insets})]. In particular, they increase linearly in $z_0$ for short distances $z_{0}/a \lesssim 1$, in agreement with the passive case and irrespective of $\vartheta_{0}$ and $\Lambda$. Increasing $\Lambda$ expedites the process for short initial distances $z_{0}/a \lesssim 1$. At large distances $z_{0}/a \gtrsim 1$ the MFPTs diverge, which occurs earlier for small~$\Lambda$. Thus, independent of the initial orientation $\vartheta_0$ resetting more frequently at short distances $z_0$ is more efficient than at large $z_0$. This can be rationalized as follows: when the time between resets becomes shorter than the time it takes the agents to reach the wall,  it becomes impossible for particles to reach the wall and thus the MFPT diverges. Fixing the resetting rate for the passive case to $\Lambda=1$ and comparing it to the active case with the same rate shows that at large $z_{0}$ the active agent is always faster ($\vartheta_{0}=\pi$, $\vartheta_{0} \sim \mathcal{U}[0,2\pi]$) or takes about an equal amount of time ($\vartheta_{0}=0$) to reach the wall. 

\begin{figure*}[tp]
\includegraphics[width =\textwidth]
{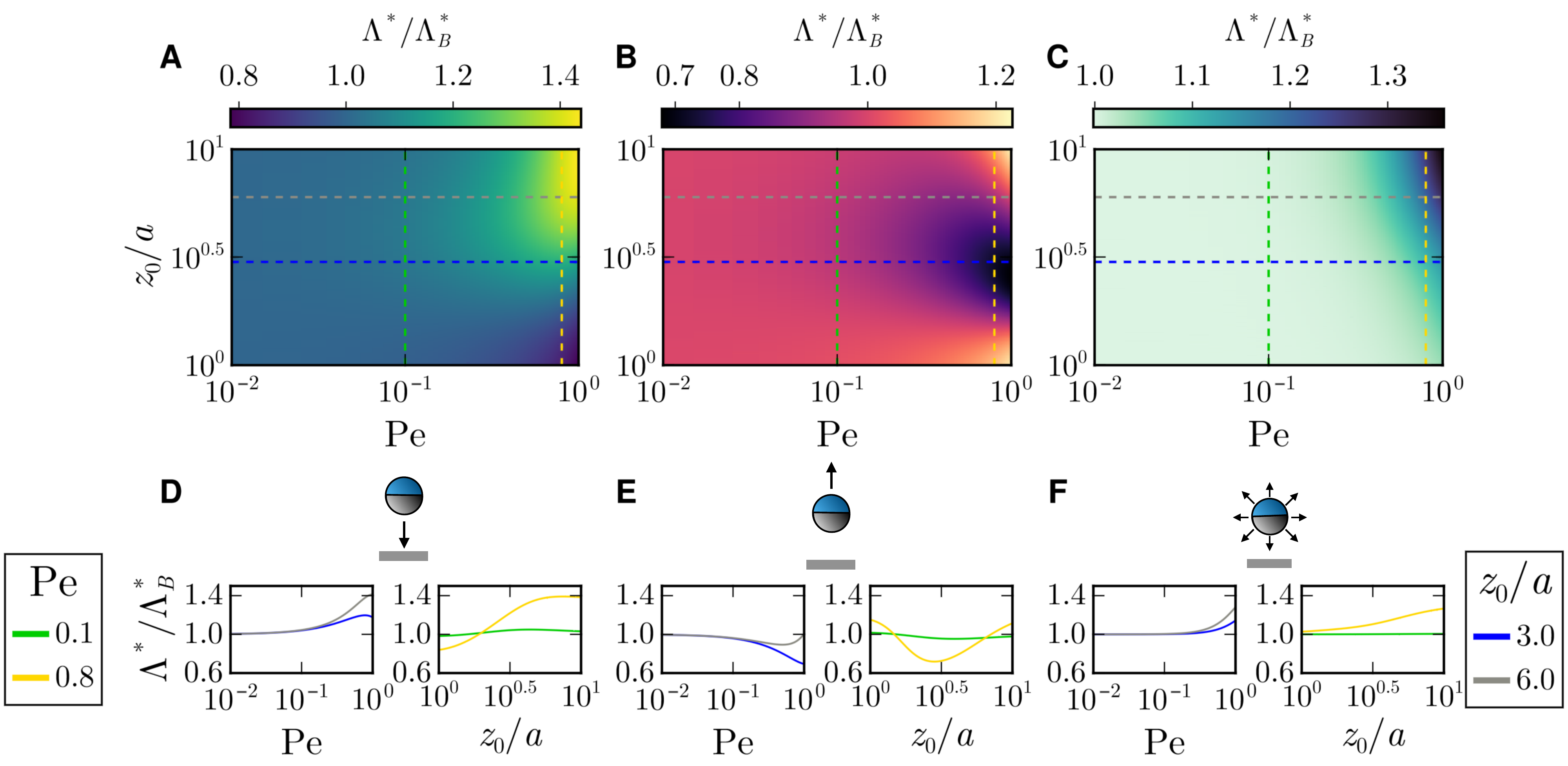}
\caption{(\textbf{A-C}) Ratio of the optimal resetting rate $\Lambda^{*}/\Lambda^{*}_{B}$ for three different initial angles $\vartheta_{0}$ as a function of the P{\'e}clet number $\mathrm{Pe}$ and of the initial position $z_{0}/a$. (\textbf{D-F}) Ratio $\Lambda^{*}/\Lambda^{*}_{B}$ as function of $\mathrm{Pe}$ (resp. $z_{0}/a$) at fixed $z_{0}/a$ (resp. $\mathrm{Pe}$). Columns show results for particles (\textbf{A,D}) initially facing the wall, (\textbf{B,E}) initially facing against the wall, and (\textbf{C,F}) with initial angles drawn from a uniform distribution $\mathcal{U}[0, 2\pi]$.\label{fig:optimum_cut}}
\end{figure*}

Our results demonstrate that the MFPTs exhibit a minimum as a function of $\Lambda$ [Figs.~\ref{fig:mfpt_optimum} (\textbf{A}-\textbf{C})], which begs the question of the optimal MFPT $\mathbb{E}[F_{t}]^{*}$ (and corresponding resetting rate $\Lambda^*$) to accelerate absorption at the boundary. We begin this discussion by comparing $\mathbb{E}[F_{t}]^{*}$ with its passive counterpart $\mathbb{E}[F_{t}]^{*}_{B}$ [Figs.~\ref{fig:mfpt_optimum} (\textbf{D}-\textbf{F})]. For low activity $\mathrm{Pe} \lesssim 0.1$, the active and passive case are comparable, $\mathbb{E}[F_{t}]^{*}/\mathbb{E}[F_{t}]^{*}_{B} \approx 1$, and the initial angle remains unimportant. Deviations appear as $\mathrm{Pe}$ increases and approaches one, where agents initially oriented towards the wall always display a lower optimal MFPT. However, agents departing away from it exhibit a slightly more complex behavior: At short initial distances $z_{0}/a  \lesssim
3$, they are naturally slower $\mathbb{E}[F_{t}]^{*}/\mathbb{E}[F_{t}]^{*}_{B} \approx 2$ because of the persistent motion, but they eventually become faster as $z_{0}/a \to 10$ as rotational diffusion kicks in and allows agents to reorient and move persistently towards the boundary. For the same reason, randomly initially oriented particles will reach the boundary faster when both activity and initial position are large enough. 

Next, we are interested in what determines this optimal resetting rate and since trying to solve analytically for $\Lambda^{*}$ leads to a lengthy transcendental equation, we rather rely on numerics and compare it with the passive case $\Lambda^{*}_{B}$ in Fig.~\ref{fig:optimum_cut}. In agreement with our observation for the optimal MFPT, the initial angle appears irrelevant for small $\mathrm{Pe} \lesssim 0.1$, thus leading to $\Lambda^{*}/\Lambda^{*}_{B}\simeq1$. 

The optimal resetting rate $\Lambda^{*}$ for a particle facing the wall ($\vartheta_{0}= \pi$) displays two behaviors depending on the initial distance: For $z_{0}/a \simeq 1$ the optimal rate $\Lambda^{*}$  decreases with P{\'e}clet number, as letting the particle reach the boundary through persistent motion is the most effective strategy. In particular, resetting as often as in the passive case would take the particle away from the boundary and increase the FPT. However, at larger initial distances, $\Lambda^{*}$ becomes larger than the passive counterpart and increases with activity. Thus, active particles need to be reset more often as they can reorient due to rotational motion, which enables them to move away from the boundary. Furthermore,  $\Lambda^{*}/\Lambda^{*}_B$ displays a maximum (see blue and gray curves in Fig.~\ref{fig:optimum_cut} \textbf{D}). The presence of this maximum can be understood by considering the length the agent moves actively before resetting $l_{R}/a=v/(a \lambda) = \mathrm{Pe}/\Lambda$. As the resetting length and the initial distance become comparable $l_{R}/z_{0} \sim 1$, resetting too often becomes disadvantageous. 

For a particle that is facing away from the boundary and for $\mathrm{Pe}$ close to one, we distinguish three cases: First, for $z_{0}/a \simeq 1$, the particle needs to be reset more frequently because of the influence of $\vartheta_{0}$. In this regime, it is likely that the particle arrives at the wall through translational diffusion, while active motion takes it away. Second, for intermediate $z_{0}/a$, resetting events need to be less frequent, to ultimately increase the chances of reorienting through rotational diffusion towards the wall and reaching it via active motion. Third and lastly, for $z_{0}/a \simeq 10$, the agent requires more frequent resets, similar to the case of a particle initially facing the boundary.

Averaging out the effect of the initial orientation $\vartheta_{0}$ shows that an active particle, as a result of persistent motion, needs to be reset more frequently at larger initial distances and for higher P{\'e}clet numbers than the passive counterpart [Fig.~\ref{fig:optimum_cut} \textbf{F}]. Interestingly $\Lambda^{*}$ seems to have no extremum when the initial angle is randomized, suggesting that resetting more frequently is always an advantage for the parameter range considered. 

\subsection{Anisotropy of the mean first-passage time\label{sec:anisotropy}}
\begin{figure}[tp]
\includegraphics[width =\columnwidth]
{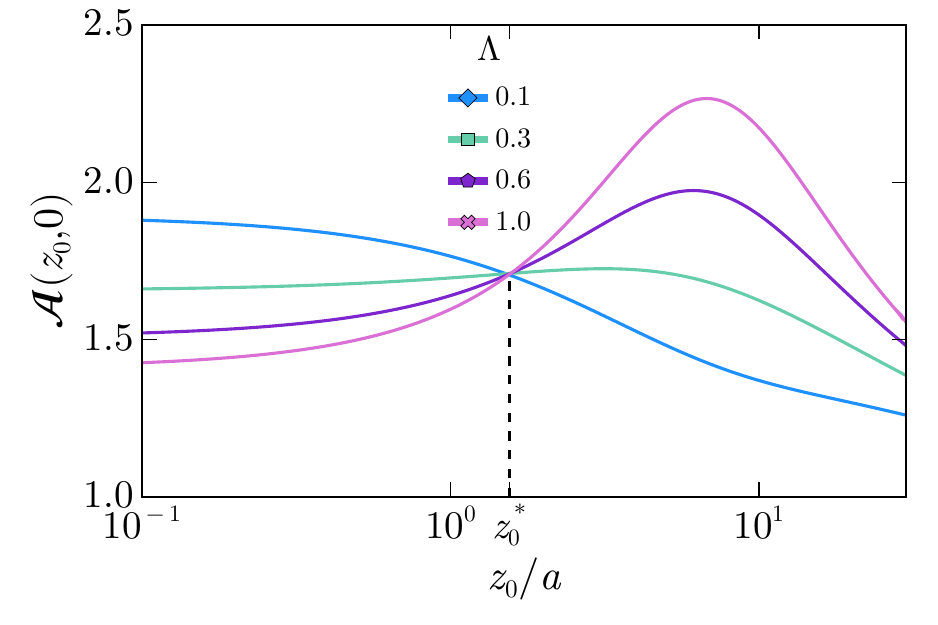}
\caption{Anisotropy $\mathcal{A}(z_{0}, 0)$  as a function of the initial position $z_{0}$ for different resetting rates $\Lambda$. The P{\'e}clet number is $\mathrm{Pe}=0.4$. \label{fig:psr_anisotropy} }
\end{figure}

To further quantify the effect of the initial orientation we introduce the anisotropy function via
\begin{equation}
    \mathcal{A}(z_{0}, \vartheta_{0}) = \frac{\mathbb{E}[F_{T}](z_{0}, \vartheta_{0})}{\mathbb{E}[F_{T}](z_{0}, \vartheta_{0} + \pi)},
\end{equation}
measuring the ratio of the MFPT given an initial position and orientation with the MFPT given the same initial position but with the diametrically opposed initial orientation. Figure~\ref{fig:psr_anisotropy} indicates that the anisotropy for an agent initially oriented towards the wall (resp. opposite to the wall) is large at small initial distances $z_{0}/a \lesssim 1$ and is a decreasing function of~$\Lambda$. This is due to the fact that at such distances, the particle initially oriented towards the wall should reach it almost immediately, which is hindered by a too frequent resetting mechanism. For $z_{0}/a \gtrsim10$, the anisotropy relaxes to $\mathcal{A}(Z_{0}, 0) \to 1$, independently of the resetting rate, as the memory of the initial orientation  progressively decays for such large initial distances. 

\begin{figure*}[tp]
\includegraphics[width =\textwidth]
{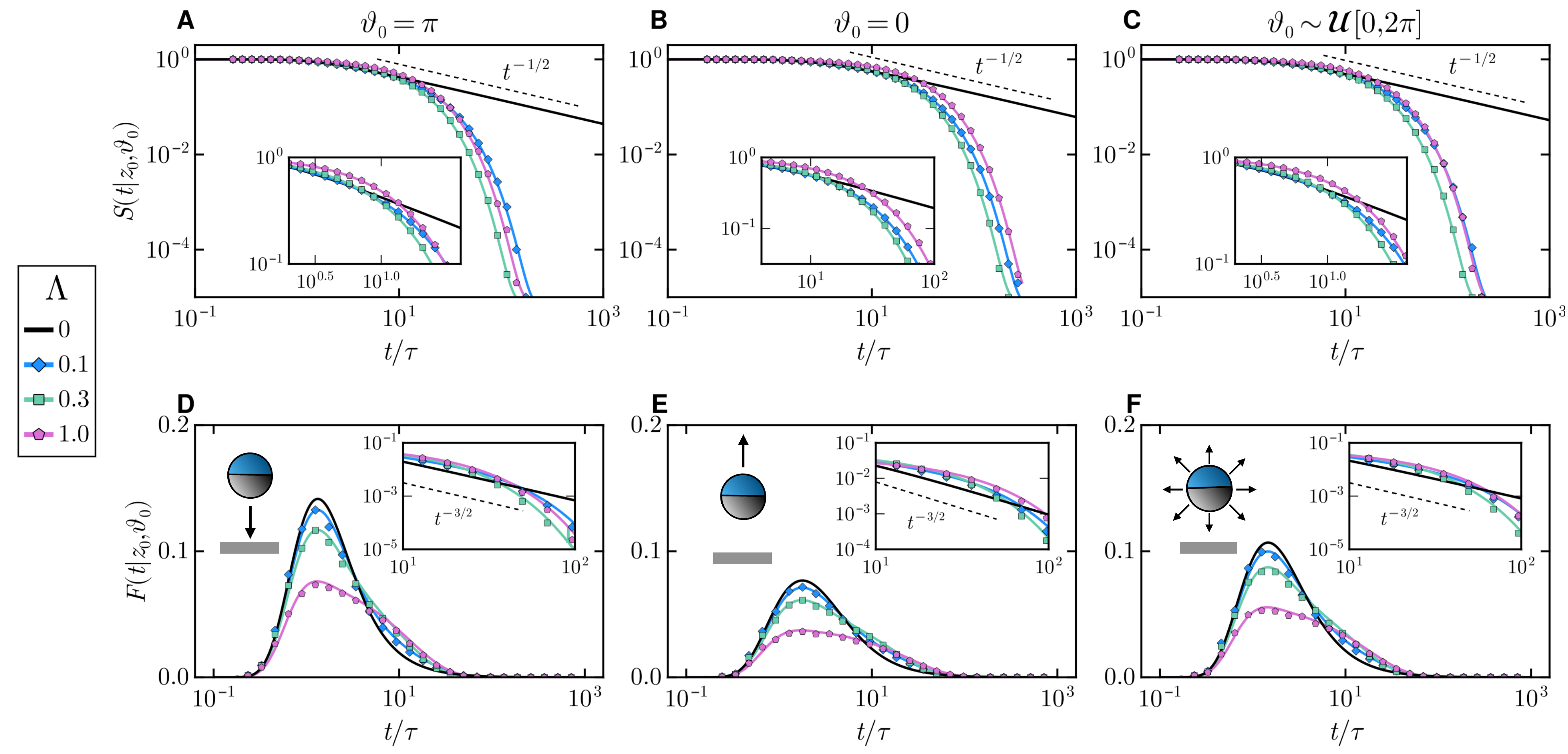}
\caption{Survival probability (\textbf{A}-\textbf{C}) and FPT probability density (\textbf{D}-\textbf{F}) as a function of time $t$ for three different initial angles~$\vartheta_{0}$: the particle is initially \textbf{A}~facing the wall, \textbf{B}~facing against the wall, and \textbf{C}~randomly oriented with an angle drawn from a uniform distribution $\mathcal{U}[0, 2\pi]$. Here, the initial position is $z_{0}/a=3$ and the P{\'e}clet number is $\mathrm{Pe}=0.4$. Solid lines and symbols denote theory and simulations for different resetting rates, respectively. The black lines represent the active case in the absence of resetting.\label{fig:fpt_psr_survival}}
\end{figure*}
At intermediate distances $z_{0}/a \simeq 1$ the anisotropy exhibits a maximum for large enough resetting rates ($\Lambda \gtrsim 0.1$) which can be rationalized in the following way: at these distances, resetting events start to be prevalent and thus effectively rectify the trajectory of agents that started moving towards the wall but oriented away and the trajectory of agents that departed away from the wall but managed to reorient towards it through rotational diffusion, strengthening the discrepancy in their MFPTs. We further note that the maximum displaces towards the right with increasing resetting rate~$\Lambda$, as the distance $l_{R} = v/\lambda = a \mathrm{Pe}/\Lambda$ traveled before resetting decreases with $\Lambda$. 

Finally, there is a point $z_{0}^{*}$ where the anisotropy appears to be independent of the resetting rate. We anticipate that it corresponds to the distance when active motion starts to become comparable to translational diffusion. Comparing the length traveled through diffusion during time $t$ ($l^{2}=t D$) with the length traveled using active motion ($l=tv$) leads to $l =a/\mathrm{Pe}$, which predicts the disappearance of that point at larger $\mathrm{Pe}$. This feature remains to be further studied for larger P{\'e}clet numbers that go beyond our perturbation limits. 

\begin{figure}[htp]
\includegraphics[width =\columnwidth]
{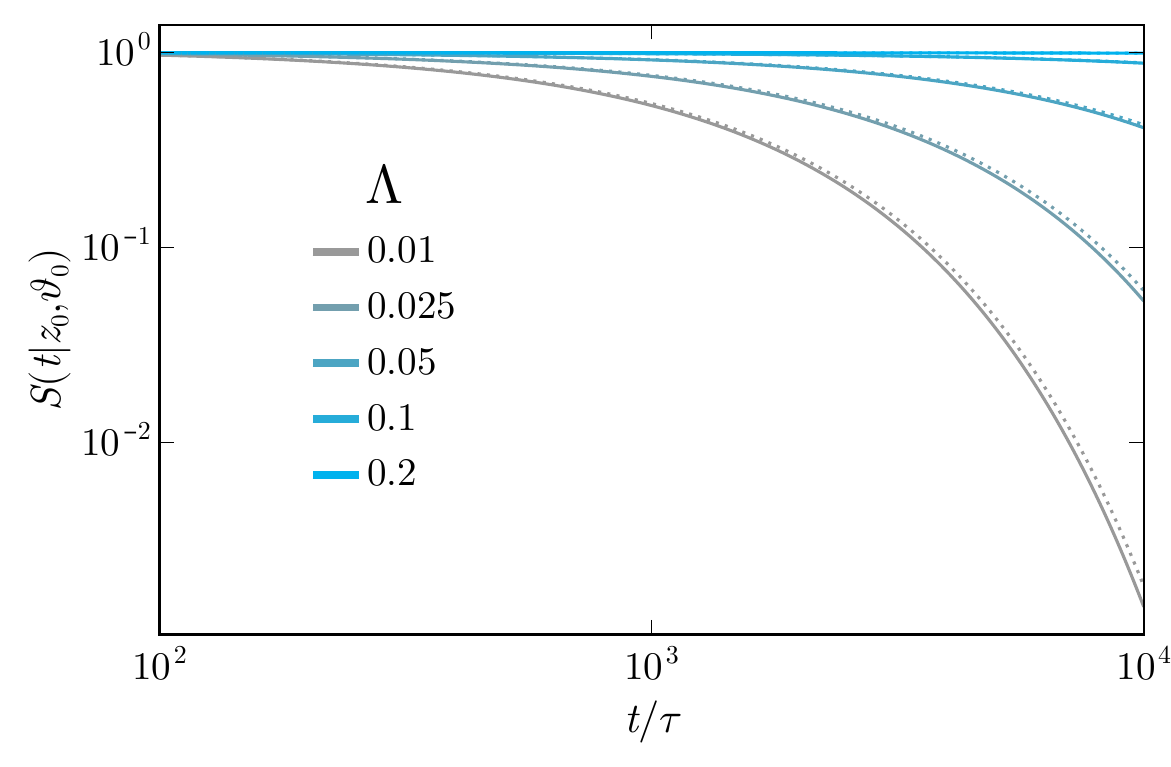}
\caption{Survival probability $S(t|z_{0}, \vartheta_{0})$ of an active particle for different resetting rates $\Lambda$. The solid lines correspond to the perturbation theory and the dotted lines represent the prediction for the effective Brownian case. Here, the initial distance is $z_{0}/a=30$, the initial orientation is $\vartheta_{0}=\pi$, and the P{\'e}clet number is $\mathrm{Pe}=0.4$. \label{fig_appendix:survival_approximation}}
\end{figure}

\subsection{Survival probability and  first-passage-time distribution \label{sec:FPTresults}}

In the past section, we studied the MFPT, which gave us a quantitative answer to the question of ``speed of completion". We can now try to deepen our understanding by also computing the survival probability $S(t|z_0, \vartheta_0)$ and the FPT probability density function $F(t |z_0\vartheta_0)$. Figure~\ref{fig:fpt_psr_survival} (\textbf{A}-\textbf{C}) shows an exponential decay for the survival probability, independently of the initial angle, and thus resetting profoundly changes the power-law tail behavior  $\sim t^{-1/2}$ of a simple ABP. The absence of this power-law tail is a sign that, unlike the non-resetting case, agents manage to reach the wall within a finite time due to the resets to their initial configuration. 

As we resolve the distributions for distances larger than the agent's persistence length, $z_{0}/a=3 \gtrsim l_{p}/a$, rotational diffusion plays an important role. Thus, the question of which survival distribution decays faster depending on $\Lambda$ follows the same logic as for the optimal MFPT: resetting is an advantage if it resists departure from the wall without hindering reorientation towards it $(\vartheta_{0}=0)$, while at the same time ensuring that there is enough time to reach it $(\vartheta_{0}=\pi)$. Furthermore, we note that the optimal resetting rate is $\Lambda^{*} \approx 0.3$ for all cases, which is reflected in the fact that the survival decays the fastest for $\Lambda=0.3$. 

We further comment on the behavior of the survival probability at large distances $z_0/a\gg l_p$. Given that at large times $t \gtrsim 1/D_{\mathrm{rot}}$, an ABP enters an effective diffusive regime characterized by the effective diffusion coefficient
\begin{equation}
    D_{\mathrm{eff}} = D \left(1+\frac{2}{3}\mathrm{Pe}^{2} \right),
\end{equation}
we suggest the survival probability of the active particle  assumes the form of that of a Brownian particle in Eq.~\eqref{eq:survival_Brownian}
by replacing the translational diffusivity by the effective diffusivity, $D=D_{\mathrm{eff}}$. Indeed, we observe that for $\mathrm{Pe}=0.4$, the survival probability at long times is well approximated by that of a passive agent performing effective diffusion. In our previous work~\cite{baoucheFirstpassagetimeStatisticsActive2025}, we have demonstrated that the memory of the initial angle is never actually lost and that the amplitude of the tail of the survival probability in the absence of resetting depends non-trivially on the original orientation. Introducing the resetting smoothens this effect but deviations from the passive case are naturally expected to increase with the P{\'e}clet number.

Putting in parallel the survival probability with the FPT probability density [Fig.~\ref{fig:fpt_psr_survival} (\textbf{D}-\textbf{F})] shows that a faster decay of $S$ leads to a faster-decaying tail for $F$, where again no power-law $F \propto t^{-3/2}$ is present, as in the absence of resetting. Finally, we observe that even though the resetting mechanism annihilates the tail, increasing the rate causes the distribution to flatten and spread over at intermediate times $ 0.1 \lesssim t/\tau \lesssim 100$. To expand on this observation, we  compute another statistical quantity, the skewness.

\subsection{Skewness of the distribution \label{sec:skewness}}
\begin{figure*}[tp]
\includegraphics[width =\textwidth]
{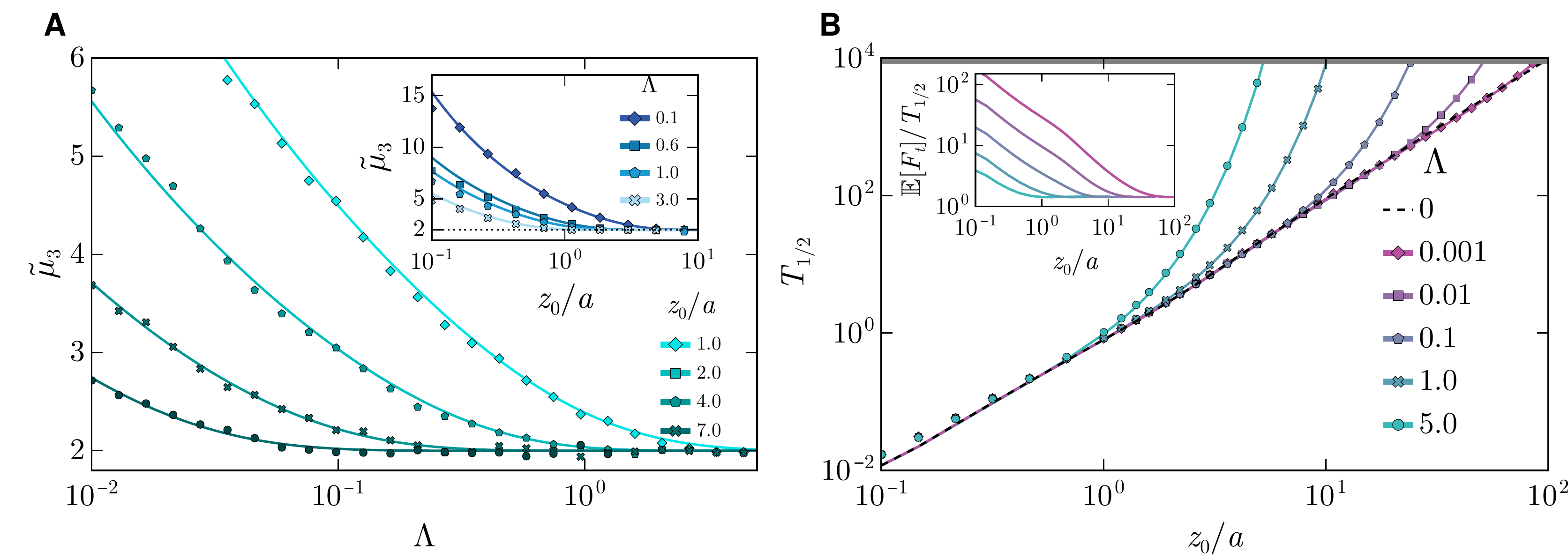}
\caption{\textbf{A}~Skewness $\tilde{\mu}_{3}$ as a function of the reduced resetting rate $\Lambda$ for different initial positions $z_{0}$ and ({\it inset}) as a function of $z_{0}$ for different $\Lambda$. The P{\'e}clet number is~$\mathrm{Pe}=0.4$ and initial orientation $\vartheta_{0}=\pi$. Theory and simulations are shown with solid lines and markers, respectively. \textbf{B}~Median $T_{1/2}$ as a function of the initial position $z_{0}$ for different reduced resetting rates $\Lambda$. The initial angle is $\vartheta_{0}=\pi$ and the P{\'e}clet number is $\mathrm{Pe}=0.4$. Theory and simulations are shown with solid lines and markers, respectively. The shaded gray corresponds to the simulation time limit. (\textit{Inset}) shows the ratio of the MFPT and the median as a function of $z_0/a$.\label{fig:skewness_median}}
\end{figure*}

Given the shape of the FPT probability density, we compute the skewness, measuring the asymmetry around the MFPT. The skewness is defined as the third standardized moment:
\begin{equation}
    \tilde{\mu}_{3} = \frac{\mathbb{E}[(F_{T}-\mathbb{E}[F_{T}])^{3}]}{\mathbb{E}[(F_{T}-\mathbb{E}[F_{T}])^{2}]^{3/2}}, \label{eq:skewness}
\end{equation}
where we use Eq.~\eqref{eq:moments_survival} to obtain the moments.
The results are summarized in Fig.~\ref{fig:skewness_median} \textbf{A} where we plot $\tilde{\mu}_{3}$ as a function of $\Lambda$ for several $z_{0}$. We first note that the skewness is positive for all cases, indicating that the FPT distribution is right-skewed, or skewed towards the large arrival times, accounting for the agents that manage to reach to wall at later times $t/\tau \gg 1$. It also diverges for vanishing $\Lambda$, in accordance with the absence of moments for the non-resetting case. For a fixed $z_{0}/a$, the skewness decreases when $\Lambda$ increases, which is consistent with the observation that the FPT probability density for $z_{0}/a=3$ stretches over a broader range of times. If the particle is increasingly reset, it naturally needs more ``tries" at the wall in order to reach it, which itself requires more time.

Interestingly, the skewness converges to 2 when the initial position is large $z_{0}/a \gg 1$ (independently of $\Lambda$) or when the resetting rate is large $\Lambda >1$ (independently of $z_{0}$). Based on our analysis of the optimal MFPT and optimal resetting rate, we expect that the skewness should converge to that of the resetting Brownian case. Using Eq.~\eqref{eq:skewness} with the leading-order term $\widehat{S}_{0}$ provides the skewness of the FPT probability density of a Brownian particle under stochastic resetting
\begin{equation}
    \tilde{\mu}_{3}^{B} = \frac{-8 + e^{\sqrt{\Lambda}Z_{0}} \left(8e^{2\sqrt{\Lambda}Z_{0}} - 3\sqrt{\Lambda}Z_{0} -12 e^{\sqrt{\Lambda}Z_{0}} \sqrt{\Lambda}Z_{0} + 3\Lambda Z_{0}^{2} \right)}{4\left(-1 +e^{2\sqrt{\Lambda}Z_{0}} - e^{\sqrt{\Lambda}Z_{0}}\sqrt{\Lambda}Z_{0} \right)^{3/2}},
\end{equation}
which converges to $2$ for $\sqrt{\Lambda}Z_0 \to \infty$, thus explaining the behavior of the skewness for the FPT probability of an active agent.

\subsection{Median of the first-passage time \label{sec:median}}
For the sake of completeness we further compute the median $T_{1/2}$, giving the value that splits the distribution into two equiprobable parts and defined as:
\begin{align}
    \int_{0}^{T_{1/2}} F(T|Z_{0}, \vartheta_{0})~\diff T = \frac{1}{2}.
\end{align}
In the non-resetting case, where the MFPT is not defined, the median allowed us to give a quantitative answer to the question of the FPT~\cite{baoucheFirstpassagetimeStatisticsActive2025}, which we compare here to the resetting case. Figure~\ref{fig:skewness_median} \textbf{B} displays the median $T_{1/2}$ as a function of the initial position $z_{0}$ for several rates $\Lambda$ and for $\vartheta_{0} =\pi$. In the case of no resetting and very low resetting rate $\Lambda \ll 1$, the median grows quadratically with the initial distance $T_{1/2} \propto z_{0}^{2}$ (as the diffusive case). For increasing $\Lambda$, $T_{1/2}$ starts to diverge at $z_{0}/a \gtrsim 10$, as the positional resetting events are too frequent for the agent to ever reach the wall. 

To further quantify the (a)symmetry of the distribution, we compare the ratio of the mean and the median in Fig.~\ref{fig:skewness_median} \textbf{A} (\textit{inset}). At small initial distances $z_{0}/a \lesssim 1 $, we first notice that for vanishing $\Lambda$, the ratio $\mathbb{E}[F_{t}]/T_{1/2}$ diverges since the MFPT itself diverges while the median remains defined. As the resetting rate increases, $\mathbb{E}[F_{t}]/T_{1/2}$ decreases due to the resetting process stabilizing the distribution. We finally note that even though both the median and the MFPT diverge at large $z_{0}/a$, their ratio seems to eventually converge to a value slightly above one, independently of the resetting rate, which is consistent with the positive skewness studied in the previous section. Thus, this further emphasizes that the distribution remains asymmetric for any parameters considered here. (Let us finally note that the results remain qualitatively similar for different initial orientations.)

\section{Conclusion \label{sec:summary}}

Here, we have studied the FPT properties of an ABP under stochastic resetting to its initial configuration. Employing a previously developed perturbative approach and a renewal framework, we compute exact expressions for the FPT probability density and several other statistical indicators such as the mean, skewness, and median. The main difference to the bare diffusive case is the additional initial orientation of the particle relative to the wall, which can make reaching the boundary slower or faster than a diffusive particle. By quantifying the optimal resetting rate our results demonstrate that active agents, which are relatively far away from the boundary, should be reset more frequently than passive ones to minimize their arrival time, as active motion can take them further away. We further discuss the effect of the initial orientation through an anisotropy function which becomes most pronounced when the initial distance is comparable to the distance the particle travels before it is reset.  

The theory developed in this work is valid for small  P{\'e}clet numbers and it is expected that the interplay of all processes changes for high activity. For instance, an agent departing away from the wall and swimming persistently at high P{\'e}clet number is unlikely to reach the wall and it would be interesting to see what the resetting mechanism changes in this case. While we have here integrated out the direction parallel to the wall, it would be interesting to explore where the particles hit the wall.

Since the agent considered here is active, there are even more directions in which one can extend this work. Experimental realizations of positional stochastic resetting for passive particles have been achieved through the implementation of a potential~\cite{guptaStochasticResettingVery2022a} and have yielded validation of the underlying theory. This is particularly relevant for our case, where the question of resetting a microscale active agent is experimentally not straightforward~\cite{abdoliStochasticResettingActive2021a}. Resetting is naturally associated with a thermodynamic cost~\cite{fuchsStochasticThermodynamicsResetting2016a,busielloEntropyProductionSystems2020,sunilCostStochasticResetting2023a} and as these active agents are generally immersed in a fluid they would also need to overcome an additional drag force. Further experimental efforts may focus on designing efficient resetting protocols for microswimmers, guided by our theoretical predictions.

In the context of foraging and other target-search problems, stochastic resetting is often depicted as an advantageous search strategy~\cite{bressloffSearchProcessesStochastic2020, guptaStochasticResettingVery2022a} but in a lot of situations agents are active and cannot be said to simply perform Brownian motion. We therefore believe that our work is relevant for establishing a thorough physical understanding of the underlying physics.

\begin{acknowledgments} 
We gratefully acknowledge discussions with Thomas Franosch and Magali Le Goff.
\end{acknowledgments}

\section*{Conflicts of interest}
There are no conflicts to declare.

\section*{Data availability}
Most data can be generated by using the formulas provided in the manuscript. The data that support the findings of this study are available upon reasonable request from the authors.

\section*{Appendix}
The appendix contains a summary of the perturbative approach used to compute the propagator in the absence of resetting [Sec.~\ref{appendix:perturbative}], the expression of the MFPT in the presence of resetting [Sec.~\ref{appendix:results}], a validation of the perturbative approach for intermediate P{\'e}clet numbers [Sec.~\ref{appendix:validation_peclet}], and finally the details of our simulations [Sec.~\ref{app:simulations}].
\appendix{
\section{Perturbative approach }
\label{appendix:perturbative}
In this section, we provide additional details regarding the approach we used for the FPT statistics of an ABP without stochastic resetting in Ref.~\cite{baoucheFirstpassagetimeStatisticsActive2025}. Therefore, we insert the expansion in Eq.~\eqref{eq:phat} up to second order in the P{\'e}clet number
into the Laplace transform of the Fokker-Planck equation [Eq.~\eqref{eq:operator_form}] to obtain a coupled set of equations that are solved iteratively:
\begin{subequations}
    \begin{align}
        (s-\mathcal{H}_{0}) \widehat{\mathbb{P}}_{0}^{\mathrm{o}}  &=\delta(Z-Z_{0})\delta(\vartheta-\vartheta_{0}), \label{eq:G}\\
        (s-\mathcal{H}_{0}) \widehat{\mathbb{P}}_{1}^{\mathrm{o}}  &= \mathcal{V} \widehat{\mathbb{P}}_{0}^{\mathrm{o}}  \label{perturb_eq},\\
    (s-\mathcal{H}_{0}) \widehat{\mathbb{P}}_{2}^{\mathrm{o}}  &= \mathcal{V} \widehat{\mathbb{P}}_{1}^{\mathrm{o}}. \label{perturb_eq2}
    \end{align}
\end{subequations}
The contribution of order $i+1$ can be computed from the contribution of order $i$ through:
\begin{align}
    &\widehat{\mathbb{P}}_{i+1}^{\mathrm{o}}(Z,\vartheta,s | Z_{0}, \vartheta_{0}) =\label{eq:pi}\\ &\int_0^\infty   \int_{0}^{2\pi}  G(s,Z,\vartheta ,  Z', \vartheta')~ [\mathcal{V} \widehat{\mathbb{P}}_{i}^{\mathrm{o}}]( Z', \vartheta',s | Z_{0},\vartheta_{0}) \mathrm{d}Z' \mathrm{d}\vartheta', \notag
\end{align}
where $G$ denotes the Green's function solving Eq.~\eqref{eq:G} with boundary condition ~$G(s, Z=0,\vartheta, Z', \vartheta')=0$. Computing $G$ is done by first solving the equation for an unbounded domain (whose solution we denote by $G^{u}$) and subtracting the contribution of an image solution to impose the boundary condition: 
\begin{align}
           &G(s,Z, \vartheta,Z', \vartheta') \notag\\
           &= G^{u}(s,Z, \vartheta, Z', \vartheta') - G^{u}(s,Z, \vartheta, -Z', \vartheta')\notag
           \\
           &=\frac{1}{2\pi}\sum_{\ell=-\infty}^{\infty}e^{i \ell (\vartheta' -\vartheta)} \frac{1}{2 p_{\ell}} \left( e^{-p_{\ell}|Z-Z'|} -e^{-p_{\ell}|Z+Z'|}\right),
\end{align}
where $p_{\ell}^{2} = s + \gamma l^{2}$. We further note that due to the symmetry of the coefficients, $p_{\ell} = p_{-\ell}$, there is no need to change the angular part of the image solution. 

Each contribution to the survival probability is then obtained by marginalizing over the position:
\begin{equation}
    \widehat{S}_{i}^{\mathrm{o}}(s | Z_{0}, \vartheta_{0}) = \int_{0}^{\infty} \widehat{\mathbb{P}}_{i}^{\mathrm{o}}(s,Z|Z_{0}, \vartheta_{0}) ~ \mathrm{d}Z.
\end{equation}
The zeroth-order solution, corresponding to the survival probability of a  passive particle, is
\begin{align}
\widehat{S}_{0}^{\mathrm{o}}(s | Z_{0}) &=  \frac{1}{s}\left(1-e^{-\sqrt{s}Z_{0}} \right),
\label{appendix:perturb_s0}
\end{align}
and the first- and second-order corrections read:
\begin{align}
    &\widehat{S}_{1}^{\mathrm{o}}(s | Z_{0}, \vartheta_{0} ) = \frac{\cos(\vartheta_{0})}{\gamma \sqrt{s}} \left( e^{-\sqrt{s}Z_{0} } - e^{-\sqrt{s+\gamma}Z_{0}}  \right),
    \label{eq:perturb_s1}\\  
    \begin{split}
        &\widehat{S}_{2}^{\mathrm{o}}(s,| Z_{0}, \vartheta_{0}) = - \frac{e^{-Z_{0} (\sqrt{s} + \sqrt{s+\gamma} + \sqrt{s+4\gamma})}}{24 \sqrt{s} \sqrt{s+\gamma} \gamma^{2}} \Big[6 e^{Z_{0}\sqrt{s+4\gamma}} \times \\
        &\left( 2e^{\sqrt{s}Z_{0}}(s+\gamma) + e^{Z_{0}\sqrt{s+\gamma}} (-2s-2\gamma + Z_{0} \gamma \sqrt{s+\gamma}) \right) + \\ 
        & \Big( 3e^{Z_{0} (\sqrt{s+\gamma} + \sqrt{s+4\gamma})}\sqrt{s} \sqrt{s+\gamma} - 4e^{Z_{0}(\sqrt{s} + \sqrt{s+4\gamma})}(s+\gamma) \\
        &+ e^{Z_{0}(\sqrt{s}+\sqrt{s+\gamma})} (4s+4\gamma -3\sqrt{s}\sqrt{s+\gamma})  \Big) \cos(2\vartheta_{0}) \Big].  
        \label{eq:perturb_s2}
    \end{split}
\end{align}
Finally, the survival probability in Laplace space in the absence of resetting up to second order in the P{\'e}clet number is given by $\widehat{S}^{\mathrm{o}} = \widehat{S}_{0}^{\mathrm{o}} + \mathrm{Pe} \widehat{S}_{1}^{\mathrm{o}} + \mathrm{Pe}^{2} \widehat{S}_{2}^{\mathrm{o}}$. The latter is inserted into Eq.~\eqref{eq:survival_reset} in the main text to obtain the survival probability with resetting. 

\section{Analytical expression for the MFPT \label{appendix:results}}
The analytical expression of the MFPT of the ABP up to to second order in the P{\'e}clet number in Laplace space reads
\begin{equation}
        \mathbb{E}[F_{T}](Z_{0}, \vartheta_{0}) = \frac{f(Z_{0}, \vartheta_{0})}{g(Z_{0}, \vartheta_{0})}, \label{eq:MFPT}
\end{equation}
where 
\begin{widetext}
\begin{subequations}
\begin{align}
        f(Z_{0}, \vartheta_{0}) =&24\gamma^{2} -12e^{-Z_{0}\sqrt{\Lambda+\gamma}}\mathrm{Pe}^{2}\sqrt{\Lambda(\Lambda+\gamma)} 
        + \frac{1}{\sqrt{\Lambda+\gamma}}\Big[ 6e^{-\sqrt{\Lambda}Z_{0}} \Big( -4\gamma^{2} \sqrt{\Lambda+\gamma} + \mathrm{Pe}^{2} \left(2\sqrt{\Lambda}(\Lambda+\gamma) -Z_{0}\gamma \sqrt{\Lambda(\Lambda+\gamma)}  \right) \Big) \Big]\notag\\
        &+ 24\left( e^{-\sqrt{\Lambda}Z_{0}}- e^{-\sqrt{\Lambda+\gamma}Z_{0}} \right) \mathrm{Pe}\sqrt{\Lambda}\gamma\cos(\vartheta_{0}) +\mathrm{Pe}^{2}\cos(2\vartheta_{0}) \Big(-3e^{-\sqrt{\Lambda}Z_{0}}\Lambda \notag\\ 
        &+ 4e^{-Z_{0}\sqrt{\Lambda+\gamma}} 
        \sqrt{\Lambda(\Lambda+\gamma)} + e^{-Z_{0}\sqrt{\Lambda+4\gamma}} \left( 3\Lambda -4\sqrt{\Lambda(\Lambda+\gamma)}\right) \Big),\\
    g(Z_{0}, \vartheta_{0}) = &\Lambda  \Bigg(6 \Big(2e^{-Z_{0}\sqrt{\Lambda+\gamma}} \mathrm{Pe}^{2} \sqrt{\Lambda(\Lambda+\gamma)} + e^{-\sqrt{\Lambda}Z_{0}} \big(4\gamma^{2} + \mathrm{Pe}^{2}(\sqrt{\Lambda}Z_{0}\gamma -2 \sqrt{\Lambda (\Lambda+\gamma)} ) \big) \Big)  \notag\\ 
        &+ 24 \left( e^{-Z_{0}\sqrt{\gamma+\Lambda}} - e^{-\sqrt{\Lambda}Z_{0}} \right)\mathrm{Pe}\sqrt{\Lambda}\gamma \cos(\vartheta_{0}) + \mathrm{Pe}^{2} \cos(2\vartheta_{0}) \Big(3e^{-\sqrt{\Lambda}Z_{0}}\Lambda -4e^{-Z_{0} \sqrt{\Lambda+\gamma}}\sqrt{\Lambda(\Lambda+\gamma)}  \notag\\ 
        &+ e^{-Z_{0}\sqrt{\Lambda+4\gamma}} \left( 4\sqrt{\Lambda(\Lambda+\gamma)} -3\Lambda \right)  \Big) \Bigg).
\end{align}
\end{subequations}
\end{widetext}
Expansion of the full expression up to second order in the P{\'e}clet number results in 
\begin{align}
\mathbb{E}[F_{T}] = \mathbb{E}_0[F_{T}]+\mathrm{Pe} \,  \mathbb{E}_1[F_{T}]+\mathrm{Pe}^2 \,\mathbb{E}_2[F_{T}]+\mathcal{O}(\mathrm{Pe}^3) \label{eq:mfpt_perturb}
\end{align}
with $\mathbb{E}_0[F_{T}]=\mathbb{E}[F_{T}]_B$ [Eq.~\eqref{eq:MFPT_B}] and the activity-induced contributions: 
\begin{subequations}
\begin{align}
\mathbb{E}_1[F_{T}] &= \frac{e^{Z_{0}\sqrt{\Lambda}}}{\gamma \sqrt{\Lambda}}\left(e^{Z_{0} \left( \sqrt{\Lambda} - \sqrt{\gamma+\Lambda} \right) }-1 \right) \cos(\vartheta_{0}),
\end{align}
\begin{align}
&\mathbb{E}_2[F_{T}] = \frac{e^{-Z_{0} \left( 3\sqrt{\gamma + \Lambda}  + \sqrt{4\gamma + \Lambda}\right) }}{24\gamma^{2}\sqrt{\Lambda (\Lambda + \gamma)}} \Big[ -6e^{Z_{0} \left( \sqrt{\Lambda} + 2\sqrt{\gamma + \Lambda} + \sqrt{4\gamma + \Lambda} \right)} \times \notag \\
&\left( 2e^{Z_{0} \sqrt{\Lambda}} (\gamma + \Lambda) + e^{Z_{0}\sqrt{\gamma + \Lambda}} \left( -2\Lambda + \gamma \left( -2 + Z_{0} \sqrt{\gamma + \Lambda}\right) \right) \right) \notag\\ 
&+24e^{Z_{0} \left( \sqrt{\Lambda} + \sqrt{\gamma+\Lambda} + \sqrt{4\gamma + \Lambda} \right)} \left( e^{Z_{0}\sqrt{\Lambda}}-e^{Z_{0}\sqrt{\gamma + \Lambda}} \right)^{2} \cos(\vartheta_{0})^{2}  \notag \\
&+e^{Z_{0}\left( \sqrt{\Lambda} + 2\sqrt{\gamma +\Lambda} \right)} \Big( 4e^{Z_{0} \left( \sqrt{\Lambda} + \sqrt{4\gamma + \Lambda} \right)} (\gamma + \Lambda) \notag \\
&-3e^{Z_{0} \left( \sqrt{\gamma + \Lambda} + \sqrt{4\gamma + \Lambda} \right)} \sqrt{\Lambda(\gamma+\Lambda)} \notag \\
&+ e^{Z_{0} \left( \sqrt{\Lambda} + \sqrt{\gamma + \Lambda}\right)} \left(3\sqrt{\Lambda(\gamma+\Lambda)} -4\gamma -4\Lambda \right) \Big)\cos(2\vartheta_{0})  \Big] .
\end{align}
\end{subequations}

\section{Validation of the perturbation approach \label{appendix:validation_peclet}}

To check the range of validity of our perturbation approach for the computation of the MFPT, we plot $\mathbb{E}[F_{t}]$ as a function of the P{\'e}clet number, $\mathrm{Pe}$ (i.e., our perturbation parameter).
\begin{figure}[tp]
\includegraphics[width =\linewidth]
{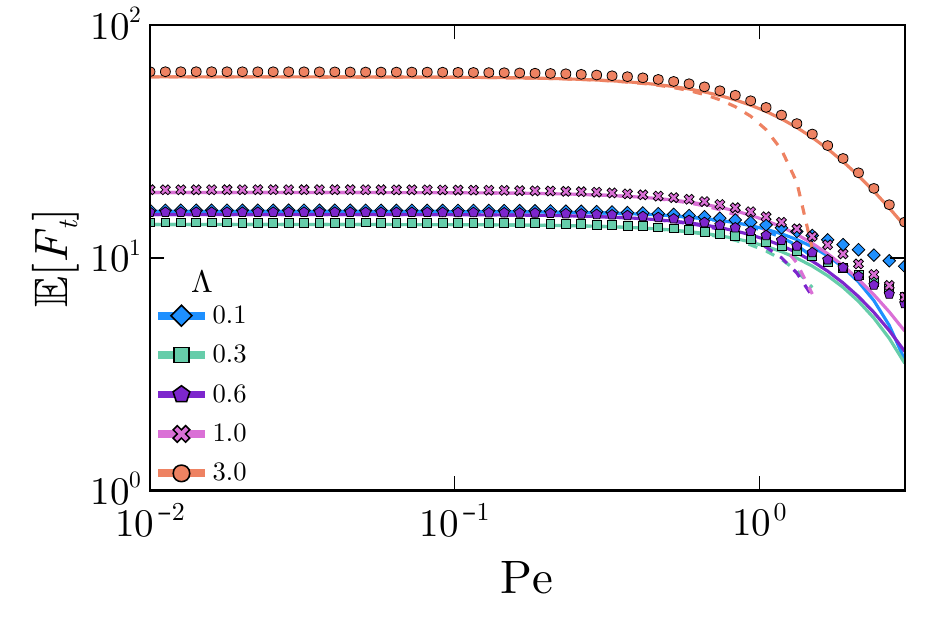}
\caption{MFPT $\mathbb{E}[F_{t}]$ as a function of the P{\'e}clet number $\mathrm{Pe}$ for several resetting rates $\Lambda$. Here, the initial distance is $z_{0}/a=3$ and the initial orientation $\vartheta_{0}$ is chosen randomly. Solid lines and symbols denote theory and simulations, respectively. Dashed lines correspond to the result given by Eq.~\eqref{eq:mfpt_perturb}. \label{fig_appendix:mfpt_pe}}
\end{figure}
Our results shown in Fig.~\ref{fig_appendix:mfpt_pe} indicate that, at least until a resetting rate $\Lambda=3$, the theory provides good agreement with simulations up to $\mathrm{Pe} \approx 1$. It also remains valid up to $\mathrm{Pe}=3$ for the largest resetting rate $\Lambda=3$, whereas discrepancies start to emerge as the resetting rate is lowered. This can be explained by considering that for $\Lambda=3$, the length traveled by the particle before resetting is smaller than the persistence length $l_{R}/l_{p}=\gamma/\Lambda \ll 1$. Thus, the agent essentially reaches the wall without fully taking advantage of persistent motion. For lower resetting rates the effects of active motion can fully develop, leading to the observed discrepancies between theory and simulations. 

In the same figure, we also plot as dashed lines the expansion of the MFPT given by Eq.~\eqref{eq:mfpt_perturb} and see that it performs worse at larger $\mathrm{Pe}$. Even though the expression given by Eq.~\eqref{eq:MFPT} is exact up to second order in the P{\'e}clet number, computing its series around zero naturally leads to truncating higher-order terms to approximate $\mathbb{E}[F_{T}]$. Our results show that the agreement with numerics becomes considerably worse at~$\mathrm{Pe}\gtrsim1$, which is why we use Eq.~\eqref{eq:mfpt_perturb} throughout the manuscript.

\section{Computer Simulations \label{app:simulations}}
To perform stochastic simulations, we discretize Eqs.~\eqref{eq:stoch_r} and~\eqref{eq:stoch_theta} according to the Euler-Maruyama scheme:
\begin{subequations}
\begin{align}
\vec{r}(t+ \Delta t) &=  \vec{r}(t) + v \vec{e}(\vartheta(t)) \Delta t + \sqrt{2D \Delta t} \vec{N}_{t}(0, 1), \\ 
\vartheta(t+ \Delta t) &= \vartheta(t) +\sqrt{2D_{\mathrm{rot}} \Delta t} N_{r}(0, 1),
\end{align}
\end{subequations}
\noindent where $\Delta t = 10^{-3} \tau$ is the time-step, $\vec{N}_{t}(0, 1)$ and $N_{r}(0, 1)$ are independent, normally-distributed random variable with zero mean and unit variance. Furthermore, the statistics are obtained by simulating trajectories for $10^5$ particles.
\bibliography{bibliography}

\end{document}